November 10, 2011

# Initial Emittance Measurements of the Fermilab Linac Beam using the MTA Beamline

C. Johnstone

## 1 Introduction

The MTA beam line has been specifically designed to facilitate measurements of the Fermilab Linac beam emittance and properties utilizing a long, 10m, element-free straight. Linac beam is extracted downstream of the 400-MeV electrostatic chopper located in the Booster injection line. This chopper cannot be utilized for MTA beam, and therefore the entire Linac beam pulse is directed into the MTA beamline. Pulse length manipulation is provided by the 750-keV electrostatic chopper at the upstream end of the Linac and, using this device, beam can be delivered from 8 µsec up to the full 50 µsec Linac pulse length.

The 10 m emittance measurement straight exploits and begins at the 12' shield wall that separates the MTA Experimental Hall and beamline stub from the Linac enclosure. A quadrupole triplet has been installed upstream of the shield wall in order to focus a large, 1.5-2" (~95% width) beam through the shield wall and onto a profile monitor located at the exit of the shielding. Another profile monitor has been installed upstream of the shield wall, ~5 m upstream of the central, or focal-point monitor. With the triplet, a small, 0.2-0.5" spot size was produced for the initial measurement reported here.

As will be shown in this report, a small, approximate beam waist located near a center profile monitor reduces the number of unknown linear optical parameters to two Courant-Snyder parameters, $\beta$ and $\varepsilon$, since, $\alpha$, or the rotation of the phase ellipse can be determined by propagating the beam envelope from this waist (using the simple linear transfer matrix that describes a drift). However, three multiwires, MW4 – MW6, with wire pitches of 2 mm, 0,5 mm, and 1mm, respectively in both transverse planes have been installed at the upstream, center and downstream positions in this straight to provide three profile measurements - with the waist approximately positioned at MW5. Three profile monitors completely determine the Courant-Snyder parameters and provide a check for assumptions of a local waist. In addition three monitors are necessary for a more detailed analysis; i.e. phase-space tomography, in the event of a non-elliptical beam. The small number of variables and the large change in beam size (with no intervening magnetic elements) reduce the systematic uncertainties and errors associated with the measurement.

Given transport beamlines are almost always dominated by linear dynamics and thus simple first-order matrix transformations, and since the invariant physical quantities being measured derive from strictly linear dynamics, background material will be presented first. Beam properties and emittance measurements are expressed in the context of these linear-dynamics approximations. (Often thin lens approximations are used and the assumption of thin lens does not alter the result). The following sections derive the analytical formulas for the emittance determination using 3 and 2 profile monitors followed by the results from the profile measurements.

## 2  Beam Properties in Linear Dynamics

Pure magnetic fields conserve energy and as such there are associated conserved quantities. In all magnetic fields both linear (dipole and quadrupole) and nonlinear ones (sextupole and higher), the total phase space area is conserved, $\sum x\,x'$, which is the sum over all the individual particles in the beam distribution. For midplane symmetric fields, the x- and y-plane phase spaces are independently conserved and this is the pure definition of the emittance of a beam. The phase space or area is not particularly insightful unless one can apply linear dynamics. In linear dynamics, fields are independent or linearly dependent on the distance of a given particle from a midplane reference trajectory (or closed-orbit in the case of a circular accelerator). Linear dynamics also require the angular or so-called kinematical term in Hamiltonian to be insignificant and not considered in order to solve the equations of motion exactly. Linear dynamics allow straightforward derivations of beam motion and most importantly, evolution of the beam envelope, in terms of simple and practical expressions involving conserved phase-space quantities. Evolution of the particle trajectory is commonly and elegantly propagated in the standard Matrix Formulation which can be used to not only track an individual particle but the beam envelope as well.

Perhaps the most powerful ansatz of linear dynamics is that as individual particle motion evolves in a linear magnetic field (or in no field), it remains constrained to an elliptical formula in phase space. The parameters of the ellipse change, but the form of the ellipse can be solved at every point along the trajectory. Knowledge of the outer particles on the ellipse at any location defines the physically-measurable extent or envelope of the beam. The evolution of the beam envelope is conventionally derived in both the sigma-matrix formalism (Figure 1) or using Courant Snyder betatron functions (also shown in Figure 1) which take advantage of an invariant physical quantity, the beam emittance, which in this case is the area of the phase-space ellipsoid. This latter is the more common definition of emittance. For linear dynamics, an elliptical phase-space area is conserved and can be systematically propagated through linear-field magnetic elements (configurations of dipoles and quadrupoles). Critical to linear dynamics and prediction is that particles remain confined to their specific ellipse, rotating around the ellipse according to a parameter called the phase advance. Particles can move in and out of the beam envelope (which is described by an ellipse which includes the majority, typically 95-99% of the particles), but do not travel inside or outside of the perimeter of their respective ellipse.

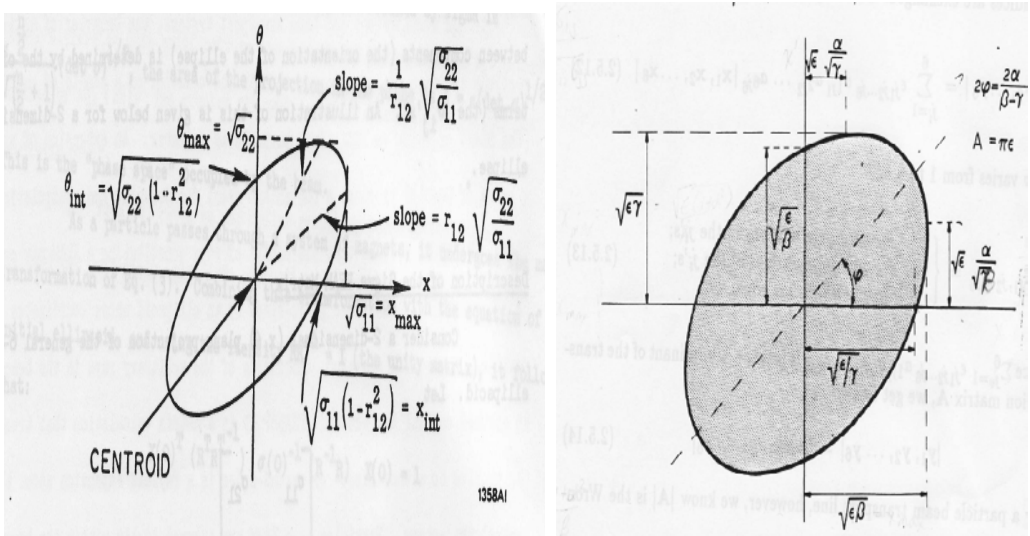

Figure 1. The linear dynamics phase space ellipse in sigma matrix notation and Courant-Snyder or accelerator notation (twiss parameters).

## 2.1 Phase Space and Emittance: Background

Given the example ellipse in Figure 1, it is clear that to measure the equation of the ellipse a priori in phase space coordinates and its area, (which is emittance in the common use of the term), a minimum of three measurements are required per plane. For the x plane, for example, x, x′ and the correlation, $r_{12}$, or tilt of the ellipse, must be determined to define an ellipse at a given location.

The extent or envelope of the beam is the physically measureable quantity and two standard formalisms are used to describe the beam envelope. The two standard implementations of the beam envelope commonly used are: 1) the sigma matrix and 2) the Courant Snyder derivation, (or so-called twiss, parameters or accelerator notation). The latter parameterization takes advantage of the invariant area of the ellipse which factors out and does not depend on position. Accelerator notation has a strong advantage that allows optical line and lattice designs without precise knowledge of the emittance or phase space area (if the phase space is even approximately elliptical).

Since both notations describe the beam envelope, their transformations are identical and the two seemingly different formalisms are related by the following simple matrix-based relationship:

$$\begin{bmatrix} \sigma_{11} & \sigma_{12} \\ \sigma_{21} & \sigma_{22} \end{bmatrix} = \epsilon \begin{bmatrix} \beta & -\alpha \\ -\alpha & \gamma \end{bmatrix} \qquad (1)$$

(Note that $\gamma$ is not actually independent variable – it is related to $\alpha$ by $(1-\alpha^2)/\beta$.) Further for a symmetric particle distribution – such as the elliptical one discussed here- $\sigma_{21} = \sigma_{12}$ so there are only three independent parameters to determine. Not also that $\alpha$ describes the correlation between position and angle, as does $r_{12}$, therefore $\alpha$ or $r_{12} = 0$ describes an upright ellipse. (In the case of an upright ellipse only two independent parameters are required to describe the phase ellipse.)

Conservation of an elliptical phase space is most easily demonstrated starting from a defined ellipse and transforming the coordinates of an individual particle on the ellipse. Let the following represent an individual particle at s=0 on an ellipse with coefficients, $\beta_0, \alpha_0, \gamma_0$:

$$\gamma_0 x_0^2 + 2\alpha_0 x_0 x_0' + \beta_0 x_0'^2 = \epsilon, \qquad (2)$$

and where the area of the ellipse is given by $\epsilon\pi$. The coordinates of the particle, $x, x'$ transform to another location, s, through the standard cosine and sine-like transformation matrix which describe the interaction trajectory of the particle with magnetic components and drifts:

$$\begin{pmatrix} x \\ x' \end{pmatrix} = \begin{bmatrix} C(s) & S(s) \\ C'(s & S'(s) \end{bmatrix} \begin{pmatrix} x_0 \\ x_0' \end{pmatrix} \qquad (3)$$

Solving Equation 2 for $x_0$ and $x_0'$ and plugging into Equation 1 gives:

$$(\gamma_0 S'^2 - 2S'C'\alpha_0 + C'^2\beta_0)x^2 + 2(-SS'\gamma_0 + \alpha_0 SC' - CC'\beta_0)xx' + (S^2\gamma_0 - 2\alpha_0 SC + C^2\beta_0)x'^2 = \epsilon. \qquad (4)$$

By redefining the parameters of the ellipse to be
$$\begin{aligned} \gamma &= C'^2\beta_0 - 2S'C'\alpha_0 + S'^2\gamma_0 \\ \alpha &= -CC'\beta_0 + (S'C + SC')\alpha_0 - SS'\gamma_0 \\ \beta &= C^2\beta_0 - 2SC\alpha_0 + S^2\gamma_0 \end{aligned} \qquad (5)$$

the equation returns to an ellipse with the same area but a different orientation and shape, but identical area. Thus an elliptical phase space is conserved in linear optics. As the transverse phase ellipse propagates through linear magnetic transport optics, it continuously changes its shape and orientation, but not its area - unless the magnetic structure is repetitive in which case both the ellipse and the beam properties (such as beam size) can repeat. Using the definitions above, the transformation matrix is given by:

$$\begin{pmatrix} \beta \\ \alpha \\ \gamma \end{pmatrix} = \begin{bmatrix} C^2 & -2SC & S^2 \\ -CC' & (S'C + SC') & -SS' \\ C'^2 & -2S'C' & S'^2 \end{bmatrix} \begin{pmatrix} \beta_0 \\ \alpha_0 \\ \gamma_0 \end{pmatrix} \qquad (6)$$

Since both the sigma matrix and the twiss functions describe the beam envelope and are related by the following matrix equality, their matrix transformations are identical

$$\begin{bmatrix} \sigma_{11} & \sigma_{21} \\ \sigma_{12} & \sigma_{22} \end{bmatrix} = \epsilon \begin{bmatrix} \beta & -\alpha \\ -\alpha & \gamma \end{bmatrix} \qquad (7)$$

Note that the emittance as defined here, the area of the ellipse, is an invariant of the equations of motion so only two parameters, $\beta$ and $\alpha$, depend on location, with $\alpha$ related to the correlation between position and angle - $\alpha = \dfrac{-r_{12}}{\sqrt{1-r_{12}^2}}$. The third parameter, $\gamma$, is somewhat superfluous, but serves to complete a straightforward and simple form for the derivation of the transformation matrix (and one which conveniently relates the Courant-Snyder parameters to the three variables in the sigma matrix description). (The sigma matrix for a symmetric particle distribution also entails only three parameters since $\sigma_{12}=\sigma_{21}$. For an arbitrary particle distribution, this is not true.)

These transformations are very important as background because they will be used in a subsequent section to derive the emittance measurement and one which uses only two profile monitors.

Ideally, fitting measured data to either the sigma matrix or the twiss functions should produce identical results. In practice, restricting the emittance to a constant when fitting measured data can prove advantageous in finding and constraining convergence to a realistic solution – hence using the Courant Snyder description can be yield more reliable results. (Often an emittance is approximately known since it is nominally defined by apertures and collimators and thus places useful limits on the search.)

### 2.1.1 Measuring the Phase Space and Emittance of a Beam

Since three parameters are necessary to describe an elliptical phase space, nominally three profile measurements are required. The most systematic-free measurement of the phase space of a beam is provided by three profile measurements at three different locations in a drift; a drift that is sufficiently long to capture a significant change in measured beam profiles. Measured profiles must be consistent with the resolution and active area of the profile monitor in order to determine accurately the profile width - the only physical parameter which is directly measured. The optimal drift is correlated to the beam emittance and the upstream focusing optics with the goal to produce a large difference in the measured profiles on different monitors. Beam divergence is the most difficult to measure and is a constant in a drift, but cannot be determined accurately for small changes in beam widths. The smaller the emittance, or the weaker the focusing in the drift, the longer the drift required to effect a measurably significant change in beam size due to smaller beam divergence.

However, if an upstream focusing system – such as a quadrupole telescope – is applied then the optics can be adjusted across a suitable drift such that a either 1) a waist or 2) a minimum spot size is established at one monitor, and then only one additional monitor is required to complete a phase space and emittance measurement. This provides additional checks on the 3-profile results – particularly in the event that data from one of the profile monitors are noisy or suspect. The following first derives the equations for an arbitrary but elliptical beam phase space as measured in a drift using 3 profile monitors and then discusses the special case in which only two profile monitors are required.

# 3  Methodology and Experimental Approach

The following describes both the full and estimated approaches to determining emittance utilizing 2 or 3 profile monitors and specifically tailored or arbitrary optics, respectively, in a straight.

## 3.1  Method 1:  Measurement using 3 Monitors in a Drift

An absolute determination of the beam phase space) can be effected by 3 profile monitors in a drift with no assumptions on beam properties outside of an invariant ellipse. The following derives the analytical solution using a Courant-Snyder parameterization of the beam envelope equation.

$$\beta_2 = C_1^2 \beta_1 - 2 C_1 S_1 \alpha_1 + \frac{S_1^2}{\beta_1}(1 + \alpha_1^2) \tag{8}$$

$$\beta_3 = C_1^2 \beta_1 - 2 C_1 S_1 \alpha_1 + \frac{S_2^2}{\beta_1}(1 + \alpha_1^2) \tag{9}$$

$$\beta = \sigma^2/\epsilon \tag{10}$$

Since ε is a constant,

$$\beta_2 = \left(\frac{\sigma_2}{\sigma_1}\right)^2 \beta_1 \tag{11}$$

$$\beta_3 = \left(\frac{\sigma_3}{\sigma_1}\right)^2 \beta_1. \tag{12}$$

Solving for $\alpha_1$ gives

$$\alpha_1 = \frac{1}{2} \beta_1 \frac{\frac{1}{S_2^2}(\sigma_3/\sigma_1)^2 - \frac{1}{S_1^2}(\sigma_2/\sigma_1)^2 - (C_2/S_2)^2 + (C_1/S_1)^2}{C_1/S_1 - C_2/S_2} = \frac{1}{2} \beta_1 \xi. \tag{13}$$

This results in

$$\beta_1 = S_1 / \sqrt{(\sigma_2/\sigma_1)^2 - C_1^2 + C_1 S_1 \xi - S_1^2 \xi^2/4} \text{ m} \tag{14}$$

and

$$\epsilon = \frac{\sigma_1^2}{S_1} \sqrt{(\sigma_2/\sigma_1)^2 - C_1^2 + C_1 S_1 \xi - S_1^2 \xi^2/4} \cdot \pi \, mm - mr. \tag{15}$$

### 3.1.1  Method 2:  Emittance determination using an optics program (MAD)

Results from the 3-monitor analytical solution can be corroborated using any optics program such as MAD and "fitting" the optics to the profile widths. The results should be identical in the case of 3 monitors. Given errors in the profile width measurements, additional monitors would further constrain the phase space solution. For the case of more than 3 monitors, a least-squares fit can be readily performed using MAD, for example.

## 3.2  Emittance measurement using 2 Monitors in a Drift

The minimum beam size in a drift always coincides with a beam waist or upright ellipse ($r_{12}$, $\alpha=0$). However, for a fixed beam profile monitor, the minimum beam size as measured at a profile monitor does not correspond to a waist; the waist occurs upstream of the minimum beam size simply because stronger focusing (a shorter focal length) produces a smaller *measured* spot size at the profile monitor - the upstream focusing lens is stronger (Figure 2). Later it will be shown that particularly for a long straight, the difference between the waist and the minimum spot size at the 2$^{nd}$ detector is insignificant insofar as the measurement of emittance is concerned. For two monitors spaced relatively equidistant from a central monitor, waist/symmetric conditions be verified by "equal" profile measurements. Although the present monitor configuration is ± 5.001/4.312 m ($L_1$ and $L_2$, respectively) about the central one, the relationship between the distance and beta function relative to a waist (assumed to coincide with the central monitor) provides a quick check of on the assumption of near-waist conditions at the central monitor.

$$\beta_1 = \beta_2 + \frac{L_1^2}{\beta_2}$$
$$\beta_3 = \beta_2 + \frac{L_2^2}{\beta_2}$$

For a tight focus at the central monitor; i.e. $\beta_2 \ll \beta_1$ or $\beta_3$.

$$\frac{\beta_1}{\beta_3} = \frac{L_1^2}{L_3^2} = \frac{\sigma_1^2}{\sigma_3^2}$$

From the distances, $L_1$= 5.001 m and $L_2$= 4.312 m, one would predict the ratio of the widths to be 1.16 and assuming a waist, an approximate $\beta_2$ can be calculated from $\sigma_2$ and $\beta_1$ and $\beta_3$ estimated, thus providing a crosscheck on the optics which will be useful in obtaining the Courant-Snyder parameters which best describe the phase space evolution of the line.

### 3.2.1  Method 3: Emittance measurement using 2 monitors in a drift: Minimum spot size at center detector

The relationship of the waist to the minimum spot size however be fully derived. This relationship is useful in that it demonstrates that only two profile monitors will still supply an accurate emittance – the errors associated with the emittance remain dominated by determination of profile widths and particularly for a linac, non-elliptical phase space.

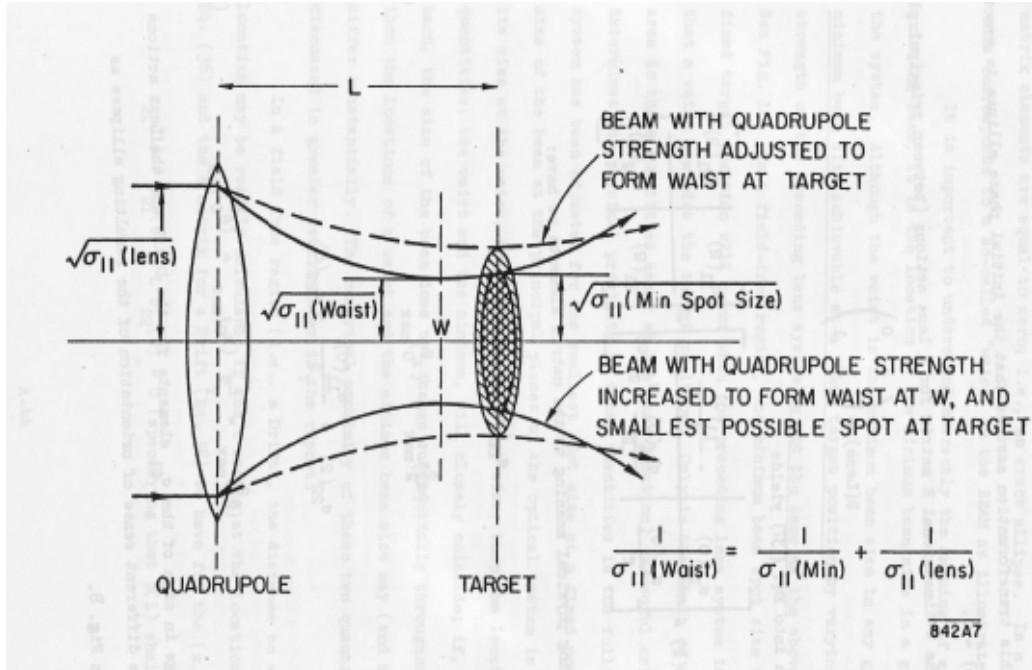

Figure 2. Picture of waist at profile monitor vs. minimum spot size

Figure 3 illustrates the focusing structure assumed in the derivations of minimum spot size and waist. The parameters characterizing the optical center of the lens system are defined to be $\beta_0$, $\alpha_0$, $\gamma_0$, and the parameters immediately after the thin-lens representation, $\beta_1$, $\alpha_1$, $\gamma_1$. Although the thin lens approximation is used, this assumption does not change the result as the net focal length is the quantity that appears in the solution.

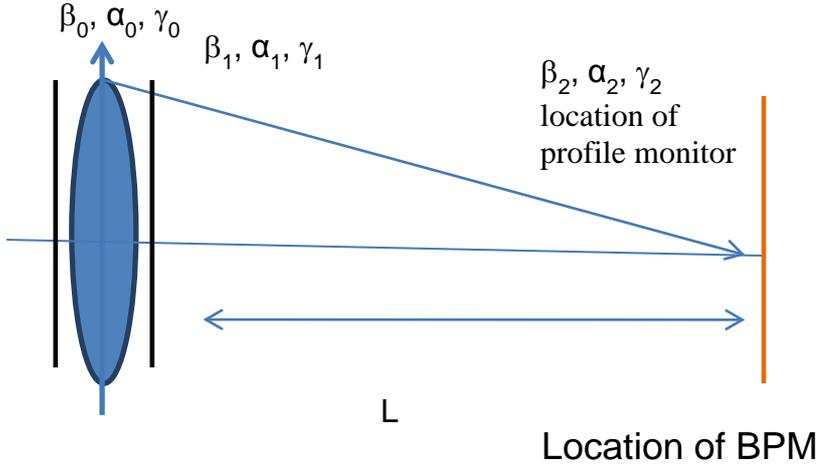

Figure 3. Focusing lens and profile monitor depicting minimum spot size optics.

The derivation of the minimum spot size optics begins with the single-particle, thin-lens transformations for a focusing structure with focal length, $f$, followed by a drift as shown in Figure 3:

$$\begin{bmatrix} 1 & 0 \\ \frac{1}{f} & 1 \end{bmatrix} \quad and \quad \begin{bmatrix} 1 & L \\ 0 & 1 \end{bmatrix} \tag{8}$$

Defining $q = 1/f$, the transformations of Equations 3 and 6 for a thin lens give

$$\begin{aligned} \beta_1 &= \beta_0 \\ \alpha_1 &= \alpha_0 - q\beta_0 \end{aligned} \tag{9}$$

Transforming to the position of the profile monitor gives:

$$\beta_2 = \beta_1 - 2\alpha_1 L + L^2 \gamma = \beta_0 - 2(\alpha_0 - q\beta_0)L + \frac{L^2(1+(\alpha_0-q\beta_0)^2)}{\beta_0} \tag{10}$$

$$\alpha_2 = \alpha_1 - L\gamma_1 = (\alpha_0 - q\beta_0) - \frac{L}{\beta_0}(1 + (\alpha_0 - q\beta_0)^2) \tag{11}$$

The minimum spot size at the position of the beam profile monitor is derived by setting the derivative of the beam envelope as a function of focusing strength to a to zero to find the minimum:

$$\frac{\partial \beta}{\partial q} = 2\beta_0 + \frac{2L^2(\alpha_0-q\beta_0)(-\beta_0)}{\beta_0} = 0 \tag{12}$$

This implies

$$(\alpha_0 - q\beta_0) = \frac{\beta_0}{L} \tag{13}$$

and therefore

$$\beta_2 = \frac{L^2}{\beta_0} \tag{14}$$

$$\alpha_2 = \frac{\beta_0}{L} - \frac{L}{\beta_0}\left(1 + \frac{\beta_0^2}{L^2}\right) = -\frac{L}{\beta_0} \quad , \tag{15}$$

which is clearly not a waist, but represents a diverging beam, so that the waist is actually located upstream when the minimum spot size is observed on the profile monitor.

Note that the condition of the minimum spot size; i.e. the zero derivative, supplies one constraint and therefore eliminates the requirement for a third profile monitor. That is, under the condition of a minimum spot size at the first monitor, an elliptical phase space is completely specified (in theory) with only one additional monitor.

One can now transform from the center monitor as shown in Figure 4 to either an upstream or a downstream monitor and solve for the optical parameters and emittance. Note that now the subscript 1 denotes the position of the upstream profile monitor (not the optical parameters just after the action of the thin-lens focusing matrix used earlier). Since only the net focal length enters into the solution, the details and physical extent of the focusing structure do not change the results – hence the simplicity of the thin lens matrices can be exploited. This latter point is particularly important because it implies this methodology is systematic-free to within the accuracy of the profile determination.

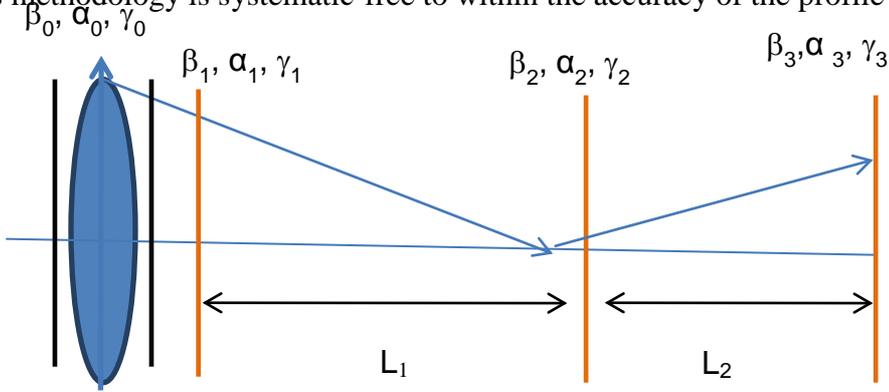

Figure 4. Three profile monitors (subscript 1, 2 and 3) in a drift showing a minimum spot size at the center monitor and an upstream waist.

The transformations to the parameters at the location of the second profile monitor are as follows:

$$\beta_3 = \beta_2 - 2\alpha_2 L_2 + L_2^2 \gamma_2 \tag{16}$$

$$\beta_2 = \frac{L_1^2}{\beta_0} \text{ and } \alpha_2 = -\frac{L_1}{\beta_0} = -\frac{\beta_2}{L_1} \tag{17}$$

Inserting the value for $\alpha_2$ gives:

$$\beta_3 = \beta_2 + 2\frac{\beta_2}{L_1}L_2 + \frac{L_2^2}{\beta_2}\left[1 + \left(\frac{\beta_2}{L_1}\right)^2\right] \tag{18}$$

The next step incorporates the measured profiles at the first and second detectors:

$$\sigma_2 = \sqrt{\beta_2 \epsilon} \text{ and } \sigma_3 = \sqrt{\beta_3 \epsilon} \tag{19}$$

where $\sigma$ is in mm, $\beta$ in meters and emittance, $\epsilon$, in mm-mr.

Substituting for $\sigma_3$ gives:

$$\sigma_3 = \sqrt{\left(\beta_2 + 2\frac{\beta_2}{L_1}L_2 + \frac{L_2^2}{\beta_2}\left(1 + \frac{\beta_2^2}{L_1^2}\right)\right)\epsilon} \quad (20)$$

$$\frac{\sigma_3^2}{\sigma_2^2} = 1 + 2\frac{L_2}{L_1} + \left(\frac{L_2}{\beta_2}\right)^2\left(1 + \frac{\beta_2^2}{L_1^2}\right) \quad (21)$$

Since the location of the profile monitors are known, the parameter $\beta_2$ can now be solved in terms of known quantities,

$$L_2^2 = \left[\frac{\sigma_3^2}{\sigma_2^2} - 1 - \frac{2L_2}{L_1} - \left(\frac{L_2}{L_1}\right)^2\right]\beta_2^2 \quad (22)$$

thus giving the emittance:

$$\varepsilon_{rms} = \frac{\sigma_2^2}{\beta_2} = \frac{\sigma_2^2}{L_2}\sqrt{\left[\frac{\sigma_3^2}{\sigma_2^2} - 1 - \frac{2L_2}{L_1} - \left(\frac{L_2}{L_1}\right)^2\right]}. \quad (23)$$

One can also use the upstream monitor and the center wire to obtain the emittance – the same analysis produces the following equation. Using the following "reverse" transformations which can be easily derived produce the following.

$$\beta_1 = \beta_2 + 2\alpha_2 L_1 + L_1^2 \gamma_2 \quad (24)$$

$$\alpha_2 = -\frac{\beta_2}{L_1} \quad (25)$$

Gives

$$\left(\frac{\sigma_1}{\sigma_2}\right)^2 = \left(\frac{L_1}{\beta_2}\right)^2 \quad (26)$$

$$\beta_2 = \frac{\sigma_2}{\sigma_1}L \quad (27)$$

$$\epsilon = \frac{\sigma_2 \sigma_1}{L_1} \quad (28)$$

This is an exact solution to within the measurement error of the profile widths and determination of the minimum spot size. However large relative differences in the profile at the two detector locations minimize this error since one is essentially measuring the divergence of the beam and its correlation to position (the $\alpha$, or $r_{12}$, term). Note that this method supplies the highest accuracy and sensitivity in measuring the divergence and the correlation coefficients as the ellipse is very proximate to upright at the second detector (it is close to the waist as will be shown later), but has a large shear or tilt to the ellipse at the beginning and the end of the drift, that is, at the location of the first and third detectors.

### 3.2.2 Method 4: Approximate emittance measurement using 2 monitors: minimum spot size at center detector

Another useful crosscheck can be performed in the case of a tight focus and produces a very reasonable estimate of the emittance. For a deep minimum (low-beta waist), the phase advance to the waist location is ~90º. In this event, the angle can be measured by the profile width at an upstream or downstream detector and the size of the beam at the focus, under the assumption it approximates a small-beta waist. A picture of the evolving phase space across a drift that depicts this useful concept is shown in Figure 5.

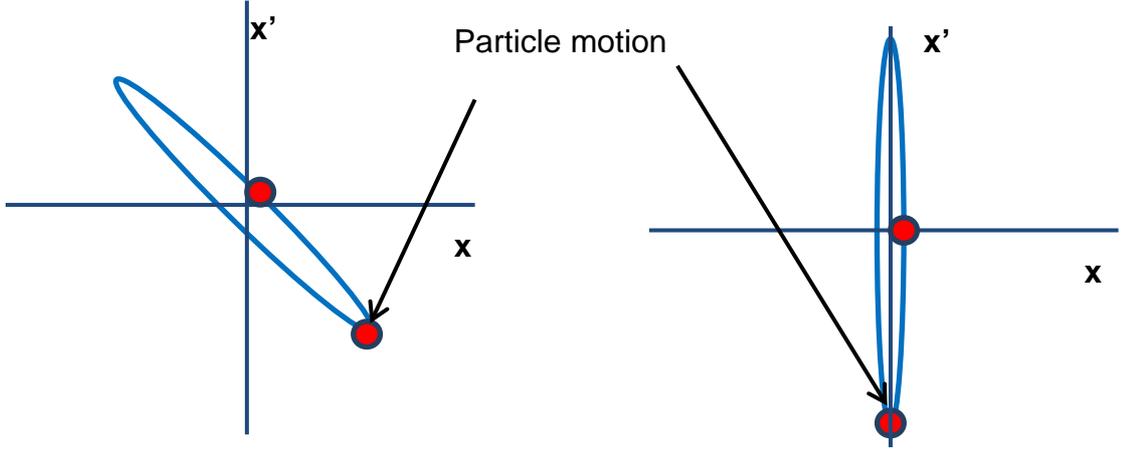

Figure 5. Evolution of an elliptical phase space from a large beta to a small-beta waist.

Therefore, since the phase-space ellipse at the detector with a small spot size (and large change in size from a distant detector) is close to a "180º" phase-advance across the waist, a very simple calculation gives a reasonable degree of accuracy for both emittance and optics. Approximating the emittance with an upright ellipse at the location of the 2$^{nd}$ detector gives the area as the minimum-size profile $x$ the beam divergence. The divergence can be measured using profiles at either the upstream or downstream detectors where the beam size is considerably larger:

$$x' or\ y' \cong \frac{\sigma_{1\,(or\,3)}}{L_{1\,(or\,2)}}.  \qquad (29)$$

At the 90º phase-advance point, or location of the waist, the emittance is given by:

$$\epsilon(x) = x \cdot x'\pi \cong \sigma_2(x) \cdot \frac{\sigma_1(x)}{L_1}\pi\ \ or\ \ \sigma_2(x) \cdot \frac{\sigma_3(x)}{L_2}\pi\ mm-mr \qquad (30)$$

$$\epsilon(y) = y \cdot y'\pi \cong \sigma_2(y) \cdot \frac{\sigma_1(y)}{L_1}\pi\ \ or\ \ \sigma_2(y) \cdot \frac{\sigma_3(x)}{L_2}\pi\ mm-mr\ . \qquad (31)$$

Depending on the measurement, the variation from the "exact" result can be on the order of 10%, which is comparable to the accuracy with which the profile width is measured. Note also that this method gives the same analytical expression as the minimum spot size and upstream/downstream method described in the previous section.

### 3.2.3 Method 5: Two Monitor emittance measurements – waist assumption

For this calculation, one exploits the equation that describes propagation of the beam envelope from a waist. Using the transformations for the beta function at a distance s from a waist, $\beta_w$, gives the standard equation for the propagation of the beam envelope from a waist:

$$\beta(s) = \beta_w + \frac{s^2}{\beta_w} . \tag{32}$$

If one uses the larger profile at the first (or third detector) and the long drift between, then to an accuracy of about a percent the beta function at the second detector reduces to:

$$\beta_1 = \beta_2 + \frac{(l+L_1)^2}{\beta_2} \cong \beta_2 + \frac{(L_1)^2}{\beta_2}; \tag{33}$$

$$\frac{\beta_1}{\beta_2} = 1 + \frac{(L_1)^2}{\beta_2} = \left(\frac{\sigma_1}{\sigma_2}\right)^2; \tag{34}$$

since $l \ll L_1$ where $l$ is the distance to the waist from MW5. Calculating the emittance gives

$$\epsilon_{rms} = \frac{\sigma_2 \sqrt{\sigma_1^2 - \sigma_2^2}}{L_1} \pi . \tag{35}$$

These values are within measurement error of the non-waist assumptions. (One can actually estimate the error by calculating the distance, $l$, to the waist.)

## 4    Data

As discussed, for the MTA beamline, a long drift with well-separated profile monitors is achieved by utilizing the 12' long shield wall effectively as the drift between the first two profile monitors, followed by a second long drift to the third profile monitor (located in the beamline stub that precedes the MTA Experimental Hall). The techniques described above can now be applied to provide measurements of Linac beam properties and emittance.

### 4.1    Profile Width Determinations

The Fermilab Linac beam deviates significantly from a Gaussian, exhibiting a more triangular shape, and therefore a weighted mean ($\mu$) and rms ($\sigma$) is calculated for channels above background ($n$=channel number) using the absolute value for the signal, $|(P(n)|: \mu = \frac{\sum_n n |P(n)|}{\sum_n |P(n)|}$ and $\sigma_{rms}^2 = \frac{\sum_n (n-\mu)^2 |P(n)|}{\sum_n |P(n)|}$. An approximate 95% (3 x rms) point is chosen as the cutoff for the channels contributing to the rms calculation. Background and noisy wires beyond the signal area cause significant error in the rms calculation. (A constant threshold is not subtracted because it makes an insignificant difference in the rms value.) In the following Figure 6, the profile wire data are depicted from MW4-6 in both the vertical and horizontal planes. Table 1 gives the results of the peak and rms values for each distribution. Note that MW6 horizontal is off center so utilizing the two-monitor profile technique is important to check the 3-monitor results for this plane.

One of the first effects observed is that the waist is offset in the vertical from profile monitor 2 simply based on the ratio of the widths from monitors 1 and 3 (which should be 1.16 but is not satisfied by the data in either plane). That said the measured width, particularly given the 2 mm wire spacing on MW4 and the off-center horizontal position on MW6, is not accurate to 0.5 mm which does not validate this simple, direct comparison.

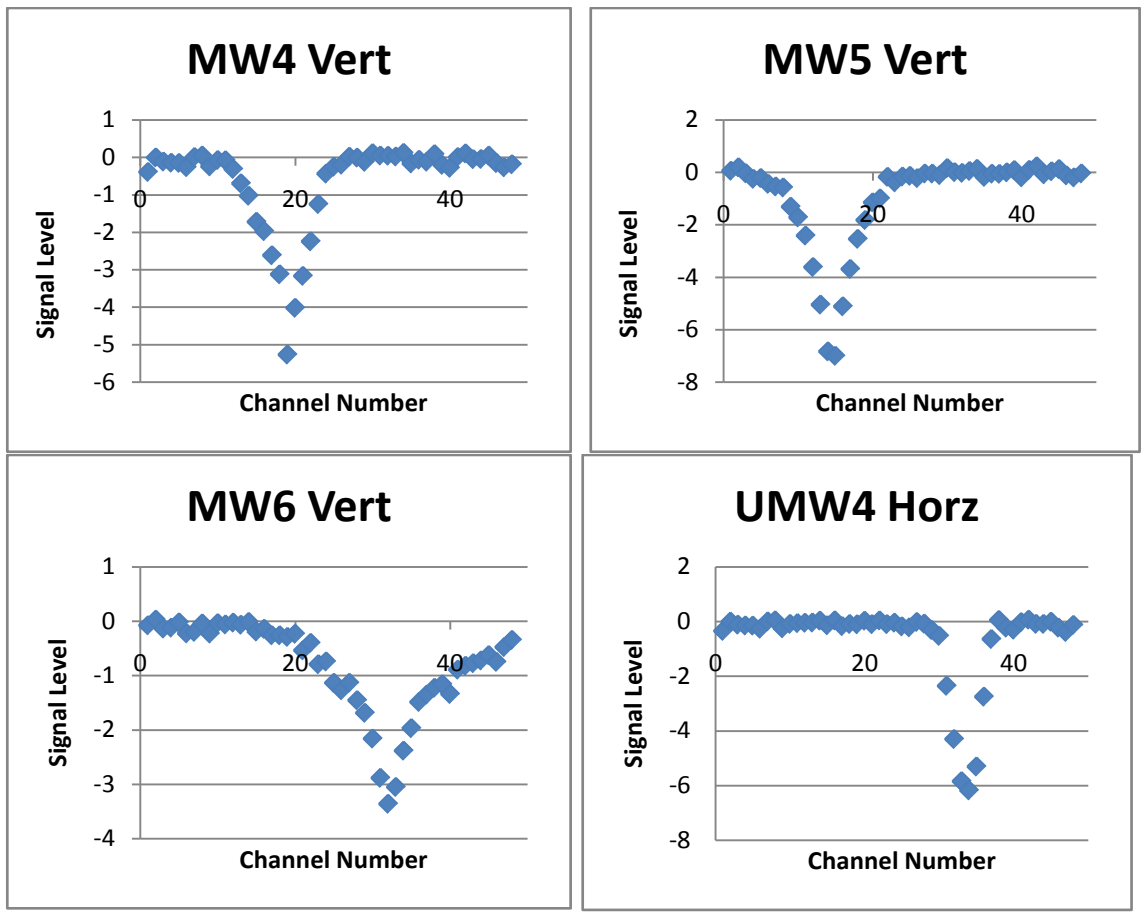

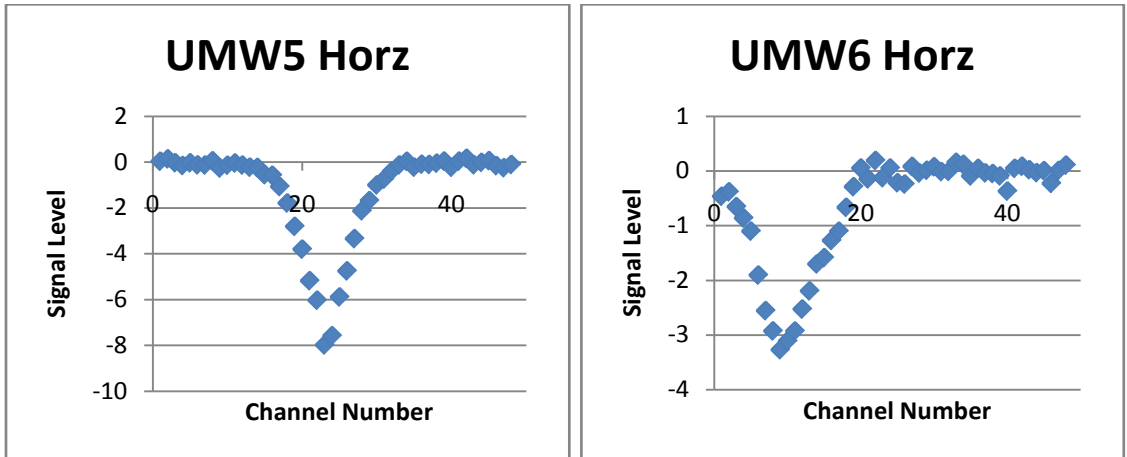

Figure 6.  Raw wire profile data plotted for MW4-6 in the MTA beamline.

Table 1.  Mean and rms values analyzing the raw wire profile data.

| Wire | Channel Range (above bkgrd) | Peak (Wire #) | RMS (# of channels) | RMS x pitch (mm) |
|---|---|---|---|---|
| Vert:  UMW4 | 6-30 | 18.58 | 3.11 | 6.23 |
| UMW5 | 1-25 | 14.52 | 3.34 | 1.67 |
| UMW6 | 10-48 | 32.80 | 6.88 | 6.88 |
| Horz: UMW4 | 25-45 | 33.59 | 2.20 | 4.40 |
| UMW5 | 15-35 | 23.43 | 3.36 | 1.68 |
| UMW6 | 1-30 | 10.32 | 4.14 | 4.14 |

## 4.2  Emittance Measurement Results

The following Table 2 summarizes the results using the 5 approaches to calculating or estimating emittance. The small difference between the 3-monitor analytical result compared with the MAD fit is simply the accuracy (decimal place) of the profile width used as input to MAD. Further error results from the horizontal profile being far off center on MW6, which produces an evident discrepancy when compared with the other approaches.

Table 2.  Results for emittance calculations using the various methods.

| Emittance | Method 1 $\pi$ mm-mr | Method 2 (MAD fit) | Method 3 Wires 1&2 (wires 2&3) | Method 4 Wires 1&2 (wires 2&3) | Method 5 Wires 1&2 |
|---|---|---|---|---|---|
| $\varepsilon_y$ | 2.00 | 1.98 | 2.08 (2.37) | 2.08 (2.66) | 2.00 |
| $\varepsilon_x$ | 1.45 | 1.39 | 1.48 (1.05) | 1.48 (1.61) | 1.37 |

Derivation of the corresponding Courant-Snyder parameters provides additional information about the waist assumption or proximity to the waist. The different methods yield the following Courant Snyder functions.

## 4.3 Results

Method 4 yields a very good estimate of the expected emittance and serves as a reliable crosscheck on the other data. It also highlights potential problems with MW6 data both vertically and horizontally due to off-center profiles. Method 3 also supports issues with using wire 3. The waist and minimum spot size approaches, Method 5 and 6, respectively yield consistent results establishing the validity of assuming these conditions and confirming that the minimum spot size and waist optics for a long straight are essentially identical.

The Courant-Snyder parameters in Table 3 also yield consistent results between the various methods. Further, the derived alphas at MW5 also indicate the waist is slightly upstream, as expected.

Table 3. Derived Courant-Snyder functions for the different methods.

| Courant Synder functions | Method 1 | Method 2 (MAD fit) | Method 3 Wires 1&2 | Method 4 Wires 1&2 (wires 2&3) | Method 5 Wires 1&2 |
|---|---|---|---|---|---|
| MW4: $\beta_y$, $\alpha_y$ | 19.34 m, 4.14 | 19.58 m, 4.13 | 18.65 m, 4.27 | 18.65 m (14.56 m) -- | 19.36 m, 3.59 |
| $\beta_x$, $\alpha_x$ | 13.33 m, 2.36 | 13.93 m, 2.43 | 13.10 3.07 | 1.48 m (1.61 m) -- | 14.18 m, 2.42 |
| MW5: $\beta_y$, $\alpha_y$ | 1.39 m -0.55 | 1.41 m, -0.54 | 1.34 m -0.27 | 1.34 m (1.05 m) -- | 1.39 m 0 |
| $\beta_x$, $\alpha_x$ | 1.94 m -0.11 | 2.03 m, -0.09 | 1.91 m, -0.38 | 1.91 m (1.75 m) -- | 2.07 m 0 |
| MW6: $\beta_y$, $\alpha_y$ | 23.59 m -4.60 | 23.88 m -4.58 | 22.76 m -3.72 | 22.76 m (17.76 m) -- | 23.62 m -3.10 |
| $\beta_x$, $\alpha_x$ | 11.78 m -2.35 | 12.33 m -2.26 | 11.58 m -2.97 | 11.58 m (10.62 m) -- | 12.53 m -2.09 |

# 5 Conclusions and Future Work

These data represent a first-pass measurement of the Linac emittance based on various techniques. No explicit tuning or centering of the beam was performed. It is clear that the most accurate representation of the emittance is given by the 3-profile approach. Future work will entail minimizing the beam spot size on MW5 to test and possibly improve the accuracy of the 2-profile approach. The 95% emittance is ~18π in the vertical and ~13π in the horizontal, which is especially larger than anticipated – 8-10π was expected. One possible explanation is that the entire Linac pulse is extracted into the MTA beamline and during the first few microseconds, the feed forward and RF regulation are not stable. This may result in a larger net emittance observed versus beam injected into Booster, where the leading part of the Linac beam pulse is chopped. Future studies will clearly entail a measurement of the emittance vs. pulse length.

One additional concern is that the Linac phase space is most likely aperture-defined and non-elliptical in nature. Unless carefully placed, rectangular apertures clip the phase space into a non-elliptical form which would not follow the analysis presented here. A non-elliptical phase-space determination would require a more elaborate analysis as discussed below. A non-elliptical phase space would might also provide another explanation of the large emittance measured.

### 5.1.1 Non-elliptical Phase Space

Measuring the Linac-beam optical functions and emittance are subject to the following caveat: the models and empirical fits assume linear dynamics and a first-order transport matrix description. An invariant elliptical phase space is therefore implicit in the propagation and prediction of beam characteristics; i.e. the beam envelope or size measured at monitored locations. If not elliptical, individual particles can be tracked (using a code such as TURTLE) and an envelope reconstructed (this methodology does not easily lend itself to design, control and optimization of transport optics and beam envelope; it is a slow iterative process with no assured convergence hence the reliance on producing elliptical phase space distributions). An arbitrary phase space distribution generally requires a specialized tomography section: at a minimum three profile monitors and the ability to "rotate" the particle distribution in phase space, with the accuracy of the reconstruction determined by the number of phase-space points measured. Figure 7 illustrates an aperture defined phase space and indicates that the envelope does not follow the conventional wisdom of either the sigma matrix or the beta function. This is generally called "tumbling" as the beam distribution is not distributed uniformly about the ellipse and, as particles on different (but concentric) ellipses rotate in and out of the beam envelope, the envelope is not predicted by propagation of the sigma matrix nor the beta functions. The following Figure 7 first illustrates a non-elliptical phase space created by an aperture and then at a location downstream with identical envelope functions. However, since the phase space is not elliptical, the extent of the measured beam profile is different at the two locations because the particles that determine the envelope are confined to different ellipses.

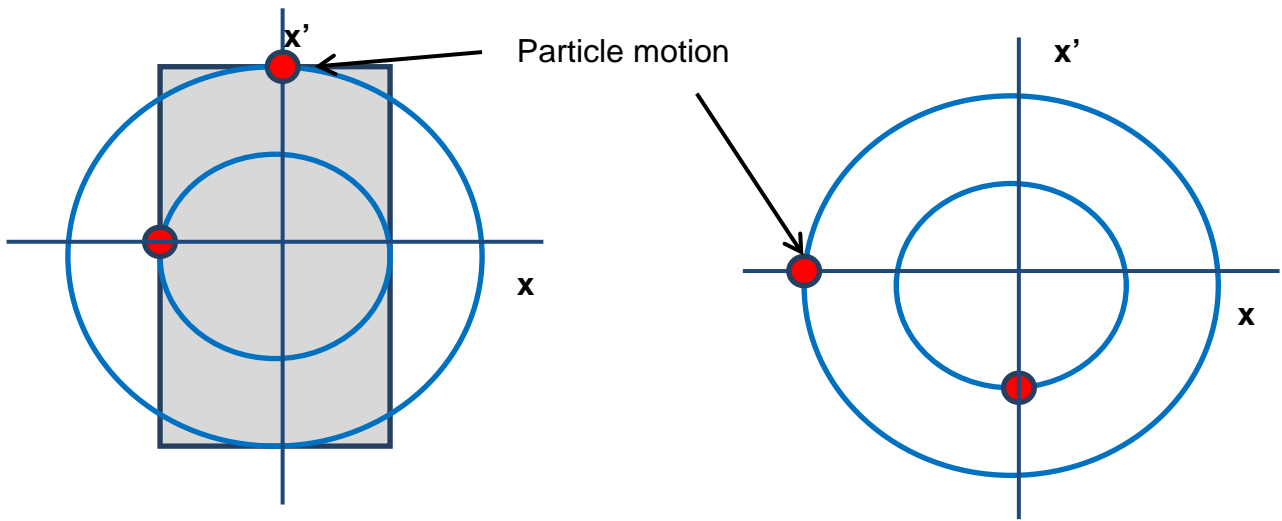

Figure 7. Initial distribution defined by an aperture (left) and a location with identical betatron functions, but π/2 downstream in phase advance (right).

Although not derived here, the phase advance of a single particle trajectory is given by

$$d\varphi \ (rad) = \int_0^s \frac{ds}{\beta}. \qquad (36)$$

Note that this advance angle is measured relative to the phase-space coordinate system not the axes of the ellipsoid. What can be established from the above equation is that the rotation of the particles in phase space is determined by the value of the "low-beta" function at the waist (MW5 in this case). The value of beta at the waist, and therefore the phase advance, can be readily varied by changing the strengths in the quadrupole triplet; thus the profile monitors can sample the phase space distribution at different phase advances or stages in their rotation. A phase space distribution (tomography) can be reconstructed from profile measurements based on different waist optics or phase space slices. A full phase space tomography is planned using this beamline.